\documentclass[nofootinbib,letterpaper,aps,amsmath,amsfonts,twocolumn]{revtex4}
\begin{document}
\newcommand{\beeq}{\begin{equation}}
\newcommand{\eneq}{\end{equation}}
\newcommand{\beeqar}{\begin{eqnarray}}
\newcommand{\eneqar}{\end{eqnarray}}
%  @twocolumnfalse\endcsname
\title{ Unknown single oscillator coherent states
do have statistical significance.
}
\author{N.D. Hari Dass\ddag} 
\address{Centre for High Energy Physics, Indian Institute of Science, Bangalore 560012, 
India}
\begin{abstract}
It is shown, contrary to popular belief, that {\it single unknown} oscillator coherent states can be endowed
with a {\em measurable statistical significance}. 
\end{abstract}
\maketitle
\vskip 2pc
%\vskip 2pc] % end \twocolumn[...]
%\vskip 2pc] 
%\pacs{10.,11.,11.10.-z,11.30.Rd,12.15.-y,12.38.Qk,12.38.-t,12.39.Fe,
%12.40.-y,12.60.-i,12.60.Cn}
%%%%%%%%%%%%%%%% Latex File %%%%%%%%%%%%%%%%%
\section{Introduction}
{\label{intro}}
Nature of single quantum states has been the subject
of a lot of debate right from the early days of
Quantum theory and has a very important role in the
interpretation of quantum theory. According to the
currently accepted picture, an {unknown single quantum
state} can not be endowed with any {\it statistical
significance} and as per the {\em ensemble interpretation } of quantum mechanics is not physically
meaningful. Though proposals like {\em Protective Measurements}\cite{protect} claiming to provide a non-destructive
measurement on certain classes of unknown single states and yet providing full information about the state
have been made, a more careful examination \cite{unprotect} has revealed that even very mild departures from Adiabaticity,
as indeed would be in any realistic measurement process, upsets the {\em Protection} in an uncontrollable
manner. See however \cite{pracprot} for a pragmatic use of the concept of Protective Measurements.

In fact this interpretation of quantum mechanics asserts
that only {\em ensembles of identically prepared states}
are meaningful in quantum theory in the sense that
only measurements on such ensembles yield unambiguous
results. This has to do with the way the act of
measurement is interpreted in quantum theory. According
to this, if an {\em observable} ${\cal A}$ is measured in any
state which is not its {\em eigenstate}, the outcome
of any {\em single measurement} is a {\em purely random
} choice of the {\em eigenvalues} of ${\cal A}$. 
Further, the state after the measurement becomes the 
corresponding eigenstate irrespective of the original 
state.

It is clear that such an interpretation of the act
of measurement introduces a dramatic difference between
the cases where the measurement is done on 
{\em unknown} and {\em known} states. When the original 
state is known, one has the {\em option} of making
the measurement of a suitable observable of which the
known state is an eigenstate. Such a measurement has
the following features: i) it {\em unambiguously}
yields the eigenvalue of the observable being measured,
ii) it does not alter the original state and iii) it
allows all compatible (mutually commuting) observables
to also be measured without any ambiguity.

In contrast, if the original state is {\em unknown},
generically any observable chosen to be measured in
this state would not be such that the unknown state
is its eigenstate. Then as per Quantum Measurement
theory, the outcome can be any of the eigenvalues
of the observable under measurement and it would be
impossible to predict which of the eigenvalues will 
be the outcome in any given measurement. Further, since
the state after measurement changes to the 
corresponding eigenstate, the original state is 
{\it irretrievably altered}. Even if {\em accidentally}
the observable happens to be such that the original
state is its eigenstate, it will not be possible to
interpret the result of the outcome without the 
explicit knowledge about this. Since the state has been
changed after the measurement, repeated measurements
subsequently  yield no {\it information} about the
original state.

It is of course possible to give a {\it Bayesian
estimate} for the 
unknown state based on the outcome of the single measurement, but there is no way of either 
confirming or improving this estimate with subsequent
measurements as the state after the first measurement
is not {\em correlated} with the initial unknown
state.

The well known {\it no cloning theorem } \cite{noclone} which asserts
that in quantum theory it is {\em impossible} to
make copies of an {\em unknown} single state, in
fact provides a remarkable consistency to the
abovementioned interpretation. For, if such a copying
were possible, one could have created an 
{\it arbitrarily large} ensemble and determined the
statistical significance through {\em ensemble
measurements}. It is often stated that {\em orthogonal}
states can be copied but it should be stressed that
even that is possible only {\em when the orthogonal family}
is known beforehand.

Motivated by the no cloning theorem there is a vast
literature on the socalled {\em optimal cloning} 
\cite{optimal1,optimal2,optimal3}. In these
implementations, one starts with the {\em original 
unknown
state} $|\alpha\rangle$ belonging to the Hilbert space
${\cal H}_A$, a number of {\em blank states}
$|b_0\rangle,|b_1\rangle...|b_N\rangle$ belonging 
respectively to
the Hilbert spaces ${\cal H}_{B_i}$ each of which is isomorphic to ${\cal H}_A$ and a number
of {\em machine states} $|m_0\rangle,|m_1\rangle.....
|m_M\rangle$ belonging to the Hilbert space ${\cal H}_M$. The combined Hilbert space has the
structure ${\cal H}_A\otimes {\cal H}_M \otimes \prod_i
{\cal H}_{B_i}$. Then the {\it optimal cloning
transformation} ${\cal T}$ has the effect
\beeq
|\alpha\rangle\prod_0^N |b_i\rangle\prod_0^M |m_j\rangle
{\stackrel{\cal T}{\rightarrow}}
\sum_{\{i,j,k\}} d_{\{ijk\}}\prod_i |a_i\rangle \prod_j |b_j\rangle \prod_k |c_k\rangle
\eneq
in such a way that {\it all} the reduced density 
matrices ${\bf \rho}_{i_0}$ obtained by tracing over the ${\cal H}_A$
states, the machine states and all the blank states
except those belonging to ${\cal H}_{B_{i_0}}$, are all
{\it identical} and with maximum overlap with the
original unknown state $|\alpha\rangle$ i.e with the
maximum possible value of $\langle\alpha|{\bf \rho_{i_0}}|\alpha\rangle$. The reduced density matrices are {\em mixed}.

It should be noted that at any given time it is not
possible to realise more than one of the reduced matrices ${\bf \rho}_i$ as different values of $i$ require
tracing over different states. This means that we can not use these optimal clonings to get any statistical
information about the original state. In fact unitarity
precludes getting the final density matrix for all the
blank states to be of the form ${\bf \rho}\otimes{\bf
\rho}\otimes{\bf \rho}....$ etc, which is just the content of the no-cloning theorem.

In this paper we wish to show that by using the concept
of {\it information cloning} proposed by us \cite{incl}
it is indeed possible to get statistical information
about {\it single unknown coherent states} though a
certain price has to be paid for this which will be
explained later.
In the case of coherent states, complete
information about the state is contained in the complex coherency
parameter $\alpha$. Thus by information cloning what we mean is the
ability to make arbitrary number of copies of coherent states 
whose coherency parameter is $c(N)\alpha$ where $\alpha$ is the coherency
parameter of the unknown coherent state and $c(N)$ is a known constant
depending on the number of copies made.

We consider $1+N$ systems of harmonic oscillators whose
creation and annihilation operators are the set $(a,a^\dag),
(b_k,b_k^\dag)$ (where the index $k$ takes on values $1,..,N$)
satisfying the commutation relations
\beeq \label{4}
[a,a^\dag]~=~1;~~[b_j,b_k^\dag]~=~\delta_{jk};~~[a,b_k]~=~0;~~[a^\dag,b_k]~=~0
\eneq
Coherent states parametrised by a complex number are given by
\beeq \label{5}
|\alpha >~=~D(\alpha)~|0 >
\eneq
where $|0>$ is the ground state and the unitary operator $D(\alpha)$ is given by
\beeq \label{6}
D(\alpha)~=~e^{\alpha~a^\dag~-\alpha^*~a}
\eneq
Let us consider a {\it disentangled} set of coherent 
states
$|\alpha >|\beta_1 >_1|\beta_2 >_2...|\beta_N >_N$, where $\alpha$ is {\em unknown}
while $\beta_i$ are {\em known} to very high accuracy, and 
consider the action of the unitary transformation
\footnote{ The most general transformation would 
involve complex$r_j$'s. But this can be reduced to 
the present form through suitable redefinitions 
\cite{incl}}
\beeq \label{7}
U = e~~^ {~t(a^\dag\otimes\sum_j r_j b_j - a\otimes\sum_j r_j b_j^\dag)}
\eneq
By an application of the Baker-Campbell-Hausdorff 
identity and the fact that 
$U|0>|0>_1..|0>_N=|0>|0>_1..|0>_N$ it is easy
to see that the resulting state is also a disentangled 
set of coherent states expressed by
\beeq \label{8}
|\alpha^\prime>|\beta_1^\prime>_1..|\beta_N^\prime>_N~~
=~~U~~ |\alpha>|\beta_1>_1..|\beta_N>_N~~
\eneq
The initial state is
\beeq \label{13}
|I>~=~D( \alpha)~D(\beta_1)_1...D(\beta_N>_N~~
|0>|0>_1..|0>_N
\eneq
Defining
\beeq \label{14}
 a(t)~=~ U~ a~ U^\dag~~~~~~~ b_j(t)~=~ U~ b_j~ U^\dag
\eneq
one easily gets the differential equations
%\beeqar \label{15}
%{d\over dt}~\tilde a(t)~&=&~-\sum_j~r_j\tilde b_j(t)\nonumber\\
%{d\over dt}~\tilde b_j(t)~&=&~~r_j\tilde a(t)
%\eneqar
\beeq \label{16}
{d\over dt}~ a(t)~=~-\sum_j~r_j b_j(t)~~~~
{d\over dt}~ b_j(t)~=~r_j a(t)
\eneq
The solutions to these eqns are straightforward to find:
\beeqar \label{17}
 a(t)~&=&~{\cos Rt}~ a~-~\sum_j~{r_j\over R}{\sin Rt}~ b_j\nonumber\\
 b_j(t)~&=&~{r_j\over R}{\sin Rt}~  a+\sum_k~
 M_{jk}(t)~ b_k
\eneqar
where $R=\sqrt(\sum_j~r_j^2)$ and 
\beeq \label{18}
 M_{jk}~=~\delta_{jk}-{r_jr_k\over R^2}(1-{\cos Rt})
\eneq
This transformation induces a transformation on the parameters
$(\alpha,\beta_j)$ which can be represented by the
matrix $ {\cal U}$ i.e ${ \alpha}_a(t)=
{\cal U}_{ab}{ \alpha}_b$. We have introduced 
the notation 
$\alpha_a$ with $a=1,...,N+1$
such that $\alpha_1 = \alpha,\alpha_k=\beta_{k-1}
(k\geq 2)$. 
Then we have
\beeq \label{19}
{\cal U}_{1a} = \left(\begin{array}{ccccc}
                             {\cos Rt}&{r_1\over R} {\sin Rt}
			     & ..&..&{r_N\over R} {\sin Rt}
			     \end{array}\right)
\eneq
\beeq \label{20}
{\cal U}_{ab} = -{r_{a-1}\over
R}~{\sin Rt}~\delta_{b1}+(1-\delta_{b1}) M_{a-1,b-1}
\eneq
where eqn (\ref{20}) is defined for $a\geq 2$. 
Equivalently
 \begin{equation} \label{21}
         {\cal U} = \left( \begin{array}{ccccc}
 {\cos Rt}&{r_1\over R}~{\sin Rt}  &..  &..  &{r_N\over R}~{\sin Rt}\\
 -{r_1\over R}~{\sin Rt}& M_{11}&.. &.. & M_{1N}\\
 .. &.. &.. &.. &\\
 .. &.. &.. &.. &\\
 -{r_N\over R}~{\sin Rt}& M_{N1}&.. &.. & M_{NN}\\

	     \end{array} \right )
 \end{equation}

We wish to choose $\{\beta_i,r_i\}$ in such a way that
all $\beta_i(t)$ become identical as we want N identical
copies. Clearly this is possible only if $r_i = r, \beta_i = \beta$.
In that case we have
\beeq
\beta_i(t) = -{\alpha\over{\sqrt N}}~\sin Rt + \beta 
~\cos Rt
\eneq
Let us first consider the choice of ${\sin Rt}=-1$ which gives N copies 
of the state $|{\alpha\over {\sqrt N}}\rangle$. This is
what we called {\em information cloning} in \cite{incl}
as the states $|{\alpha\over{\sqrt N}}\rangle$ and
$|\alpha\rangle$ have the same information content. This particular choice of
$Rt$ will be seen to be optimal in the sense that it gives the least variance
in the estimation of $\alpha$. In this case the value of $\beta$ is immaterial.

We can now use the $N$ copies of $|{\alpha\over\sqrt N}$
to make {\it ensemble measurements} to estimate ${\alpha\over\sqrt N}$ and consequently $\alpha$. 
One can estimate a state arbitrarily accurately by
using a {\em sufficiently large} ensemble.
However, in our proposal
even though the number of copies $N$ can be
arbitrarily {\em large}, the coherency parameter given by 
${\alpha\over\sqrt N}$ becomes {\em arbitrarily small} while the
{\em uncertainties} in $\alpha$ remain the same as in the original state.
This raises the question as to how best the original state can be reconstructed
and about the resultant statistical significance 
of the unknown single coherent state.

On introducing the {\em Hermitean} momentum and position operators $\hat p,\hat x$ through
\beeq \label{43}
{\hat x} = {(a+a^{\dag})\over\sqrt 2}~~~~ {\hat p} = {(a-a^{\dag})\over\sqrt 2i}
\eneq
the {\em probability distributions } for position and 
momentum in the coherent state $|{\alpha\over\sqrt N}\rangle$ are given by
\beeqar \label{dist}
|\psi_{clone}(x)|^2 &=& {1\over\sqrt\pi} e^{-(x-\sqrt {2\over N} \alpha_R)^2}\nonumber\\ 
|\psi_{clone}(p)|^2 &=& {1\over\sqrt\pi} e^{-(p-\sqrt {2\over N} \alpha_I)^2} 
\eneqar
Let us distribute our $N$-copies into two groups of 
$N/2$ each and use one
to estimate $\alpha_R$ through position measurements 
and the other to
estimate $\alpha_I$ through momentum measurements.
Let $y_N$ denote the average value of the position
obtained in $N/2$ measurements and let $z_N$ denote the
average value of momentum also obtained in $N/2$
measurements. The {\em central limit theorem} states
that the probability distributions for $y_N,z_N$ are
given by
\beeqar \label{45}
\label{central}
f_x(y_N)&=&\sqrt{N\over 2\pi}e^{-{N\over 2}(y_N-\sqrt{2\over N}\alpha_R)^2}\nonumber\\
f_p(z_N)&=&\sqrt{N\over 2\pi}e^{-{N\over 2}(z_N-\sqrt{2\over N}\alpha_I)^2}
\eneqar
The {\em estimated} value of $\alpha$ is
\beeq \label{46}
\alpha_{est} = {\langle y_N+iz_N \rangle\over\sqrt 2}\sqrt N = \alpha
\eneq
Thus the original unknown $\alpha$ is correctly estimated. But this is not enough
and one needs to know the reliability of this estimate. For that one needs the variance.
The variances in $y_N,z_N$ given by eqn.(\ref{central}) are
\beeq
\Delta y_N = \frac{1}{\sqrt{N}}=\Delta z_N
\eneq 
resulting in the variance for $\alpha$ of
\beeq
\Delta \alpha_R = \Delta \alpha_I = \frac{1}{\sqrt{2}}
\eneq
Thus, while the statistical error in usual measurements goes as $\frac{1}{\sqrt{N}}$,
and can be made arbitrarily small by making $N$ large enough, information cloning gives
an error that is fixed and equal to the variance associated with the original unknown state.

One practical problem would be due to the fact that $N\rightarrow\infty$ the coherence parameter
for the information cloned states becomes very small leading to large noise. This can be circumvented
by making a different choice of $Rt$. For example the choice $\sin Rt = \frac{1}{\sqrt{2}}$ would
yield information cloned states with parameter $\frac{\alpha}{\sqrt{2N}} + \frac{\beta}{\sqrt{2}}$.
With a large enough $\beta$ the signal to noise ratio can be improved. further, if the errors in
the prior knowledge of $\beta$ (recall that $|\beta\rangle$ are {\em known} states) is much better than
$\frac{1}{\sqrt{N}}$, the tiny $\alpha$-dependent part can still be extracted, but the variances
will now have increased to $\Delta\alpha_R=\Delta\alpha_I=1$. There is also a way of tackling the signal
to noise problem without significantly compromising the variance. For this choose $\sin Rt$ to be very close to -1,
say $-1+\epsilon$. Then the cloned states parameter will be $\alpha (1-\epsilon)/\sqrt{N} + \sqrt{2\epsilon}\beta$.
Choosing a large enough and accurately known $\beta$ one can avoid the small signal to noise problem, but the
variance will have increased only very slightly by a factor $\frac{1}{1-\epsilon}$.

Thus we have shown that even when the coherent state is {\em unknown} single state, information cloning
will allow its determination. Of course the statistical errors can not be reduced but it is still
a long way from not being able to know anything at all about the unknown state.

\section{Acknowledgements}
The author would like to express his gratitude to the Department of Atomic Energy for the award of a
Raja Ramanna Fellowship which made this work possible, and to CHEP, IISc for its invitation to use
this Fellowship there.

%\end{references}
\end{document}